\documentclass[twocolumn,prl,superscriptaddress]{revtex4}
\usepackage{epsfig}
\usepackage{amsmath}
\usepackage{amssymb}
\usepackage{citesort}

\begin{document}

\title{A model for the accidental catalysis of
protein unfolding {\it in vivo}}

\author{Richard P.\ Sear}
\email{r.sear@surrey.ac.uk}
\affiliation{Department of Physics, University of Surrey,
Guildford, Surrey GU2 7XH, United Kingdom}





\begin{abstract}
Activated processes such as protein unfolding are
highly sensitive to heterogeneity in the environment. We study
a highly simplified model of a protein in a
random heterogeneous
environment, a model of the {\it in vivo} environment. It is found that
if the heterogeneity is sufficiently large
the total rate of the process is essentially a random variable;
this may be the cause of the species-to-species
variability in the rate of prion protein
conversion found by Deleault {\it et al.} [Nature, 425 (2003) 717].
\end{abstract}

\maketitle

Protein unfolding is implicated in a number of diseases including
prion diseases such as Creutzfeldt-Jakob disease
\cite{dobson01,harris99,aguzzi04}.
It is an activated process, a free energy barrier must be overcome
for a protein to unfold from its native state. At the top
of the barrier the protein is in the transition state for unfolding,
and the transition state's free energy
determines the rate \cite{finkelstein,creighton}.
As the rate depends exponentially on the
free energy, the rate
is very sensitive to interactions of other
molecules from the environment with the transition state.
Inside living cells there is
a mixture of thousands of different proteins,
RNA, etc., if any of them can interact with the transition state of
unfolding such that its free energy
is only a few $k_BT$ lower, then the rate
of prion protein conversion when interacting
with this other molecule will be increased by an
order of magnitude.
Supattapone and coworkers \cite{deleault03} studied
prion conversion in cell extracts and found that the
rate of prion protein conversion
was greatly accelerated by an RNA molecule or molecules, and that
surprisingly
this acceleration was specific to the RNA of only some species.
Here we look at a very simple model
of unfolding {\it in vivo}, and examine how the rate of unfolding
is affected by the protein being in a complex mixture of many other
molecules.
Characterising the interactions of thousands of different molecules
with the transition state is a hopeless task and so we resort
to a statistical approach \cite{wigner67,sear03}.
We take the interactions to be random variables.
This reduces the problem from characterising a huge number
of interactions to just characterising the distribution function
of these random variables. By taking all the interactions to be
random variables we are ignoring the fact that natural selection
may be acting to restrict or increase the strength of some of
the interactions, and so our model will be a poor one if the
RNA accelerating the rate of prion protein conversion has evolved
to interact strongly with the prion protein. Very little is
definitely known about the function of the prion protein
\cite{harris99,aguzzi04}
and so we cannot rule out this possibility.
We find that 
if the free energies of interaction with the transition state
are spread over a wide range, unfolding occurs predominantly
with the transition state in contact with one or a few of the other molecules
present.
These molecules are
the ones responsible for the outliers of the
distribution of interactions with the transition state, they are the
ones that interact most strongly with the transition state.
If we take these outliers to be RNA molecules then the predictions
of our model are consistent with the experimental findings
of Supattapone and coworkers \cite{deleault03}.
When one or a few outliers dominate the rate, it may vary significantly
from species to species simply due to
chance species-to-species variations
in the nucleotide or amino-acid sequences of these outliers.

\begin{figure}
\caption{
\lineskip 2pt
\lineskiplimit 2pt
Schematic representation of our starting model
for the transition state in contact with a patch of surface.
The surface is assumed planar for simplicity.
Hydrophobic monomers are shown as the dark cubes, 
and hydrophilic monomers are the light cubes.
The transition state is the set of $n_M=7$ contiguous monomers, $B=5$
of which are hydrophobic, on top of the surface.
\label{model}
}
\begin{center}
\epsfig{file=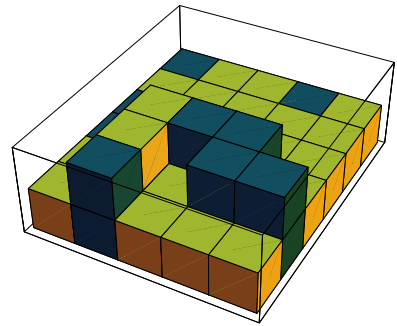,width=2.0in}
\vspace*{-0.8cm}
\end{center}
\end{figure}

Supattapone and coworkers \cite{deleault03} have shown that the
conversion of a prion protein from the PrP$^{\rm C}$ form
to the PrPres form is greatly accelerated by a specific RNA molecule
or by a small set of such molecules. The PrP$^{\rm C}$ form
is the normal form while the PrPres form is analogous
to the form associated with disease. The PrPres form is
so-called because it is Protease RESistant, i.e., not destroyed
by the proteases that cut the chains of normal proteins.
The two forms of the protein
have the same amino-acid sequence, they differ
only in conformation.
The interconversion is known to be
accelerated
by PrPres itself but Supattapone and coworkers showed that
a specific fraction of RNA molecules from both hamsters and mice but
{\em not} the same fraction from invertebrates, also appeared
to accelerate the conversion of the same protein. Of
course, in terms of the prion diseases in different species the
prion protein itself will vary from species to species and this
will cause variability. Here we are considering variability
not in the prion protein itself but in a cofactor
that interacts with the prion protein. There is
other experimental data on possible cofactors affecting the
rate of prion protein conversion.
Cordeiro {\em et al.} \cite{cordeiro01}
suggest, on the basis of
experimental evidence, that DNA reduces the free energy barrier to
the conversion of a prion protein into the form associated with
the disease.
Other work on prions has implicated as a cofactor
not RNA but a protein dubbed protein X \cite{ryou03}.
There is considerable uncertainty
surrounding the mechanism behind prion diseases \cite{harris99}.
See the reviews of Harris \cite{harris99}
and of Aguzzi and Polmenidou \cite{aguzzi04} for an introduction to prions.

We assume the unfolding of a protein to be a simple
activated process
\cite{finkelstein,creighton}, its
rate having an exponential dependence on the
barrier to unfolding, $\Delta F^*$: the difference
in free energy between the folded protein and the transition state.
The transition state being, by definition, the state of the protein
along the unfolding pathway that has the highest free energy.
Our model of the transition state for unfolding is a
linear polymer on a simple cubic lattice, $n_M$ monomers long.
Inside a living cell,
there are the surfaces of proteins, of membranes, of DNA etc..
For simplicity
we lump all these surfaces together into a large flat surface
which we model by a plane of lattice sites.
A transition state
in contact with a part of this surface
is shown in fig.~\ref{model}.
The monomers of the transition state and of the surface
are either hydrophilic or hydrophobic. We take $B$ of the monomers
of the transition state to be hydrophobic.
We assume that unfolding proceeds by some part
of the protein,
$n_M$ monomers long, unfolding, its free energy increasing
as it does so until the free energy reaches a maximum at
the transition state \cite{foldnote}. This transition state
can contact the surface, as seen in fig.~\ref{model}, and
for each hydrophobic monomer of the transition state
in contact with a hydrophobic monomer of the surface
there is a contribution of $-\epsilon$ to the free energy of the transition
state.
The only energy of interaction is between hydrophobic monomers.

The surfaces are those of proteins, RNA., etc.~and so
are coded for by the genome of the organism.
Thus they will differ between one species and another.
We have no means of calculating them from the genome of an organism
and so resort to modelling the surface with a purely random
distribution of hydrophobic and hydrophilic monomers.
Each monomer is hydrophobic with probability $h$ and
hydrophilic with probability $1-h$.
This is in the spirit pioneered by Wigner and
others \cite{wigner67} in random matrix theory, see
ref.~\cite{sear03} for an application to protein mixtures.
The surface provides $N_s$ different positions and configurations
of the transition state
in which the transition state can interact with the surface,
we call these unfolding configurations.
We neglect any correlations between the interaction
energy at different unfolding configurations on the surface and assume
that the $N_s$ configurations are independent.
Then, if we denote the 
free energy of the transition state when it is not interacting
with any other monomers by $\Delta F^*_0$, the rate of unfolding
at configuration $i$, $R_i$, is
\begin{equation}
R_i=\nu\exp\left(-\Delta F^*_0+n_i\epsilon\right),
\label{ri2}
\end{equation}
where $n_i$ is the number of hydrophobic monomers of
the transition state that are adjacent to hydrophobic parts
of the surface.
Thus, the surfaces present
are specified
by the set of $N_s$ values of the random variables $n_i$. Note that
we have assumed that the attempt frequency $\nu$
is the same for all unfolding configurations, only the free-energy barrier
varies.
We use units such that the thermal energy $k_BT=1$.

The rate of unfolding averaged over all $N_s$ possible configurations is
\begin{equation}
R=N_s^{-1}\sum_{i=1}^{N_s} R_i.
\label{hetnuc}
\end{equation}
Although we have used the specific example of protein unfolding,
quite generally the rates of activated process are given by
equations with the form of eq.~(\ref{ri2}) and so our theory
will apply quite generally to activated processes {\it in vivo}.
Equations similar to eqs. (\ref{ri2}) and (\ref{hetnuc}) were employed by
Karpov and Oxtoby \cite{karpov96} to study nucleation,
an activated process like unfolding,
in the presence of random static disorder. The author
has also applied the approach used here to nucleation \cite{sear04},
and this reference may be consulted for more details
of the analysis performed below.
The analysis required for nucleation is very similar to that required
for our model of unfolding.

Different organisms have different genomes and so different sets
of proteins etc., inside their cells. Supattapone and
coworkers \cite{deleault03} found that RNA molecules
from mammals accelerated the protein conformational conversion
whereas RNA from invertebrates did not. Thus we would like to model
and try to understand species-to-species variability. To do so
we simply assume that the surfaces in different species are uncorrelated,
then two species are modelled by two uncorrelated realisations
of the surface.
Of course, the surfaces present in closely related species in particular
will be correlated due to their similar genomes, but
we will leave the introduction of such
correlations to future work.

Continuing, as only hydrophobic monomers interact, $n_i$ is
a sum of $B$ independent random variables that are 1 with probability
$h$ and 0 with probability $1-h$.
So, the probability distribution function of $n_i$, $p(n_i)$, is
\begin{eqnarray}
p(n_i)&=&\frac{B!}{n_i!(B-n_i)!}h^{n_i}(1-h)^{B-n_i}\nonumber\\
&\simeq &\frac{\exp\left[-(n_i-m)^2/(2w^2)\right]}
{\left(2\pi w^2\right)^{1/2}},
\label{pdfni}
\end{eqnarray}
where we have indicated that $p(n_i)$ is approximately a Gaussian
for large $B$ and $n_i$.
$m=Bh$ is the mean, and the variance $w^2=Bh(1-h)$.
From now on we will neglect any deviations from
the simple Gaussian distribution function of eq.~(\ref{pdfni})
and the discrete nature of $n_i$ and
use this equation for $p(n_i)$.

Having chosen to model different species by uncorrelated realisations, we
will examine fluctuations of the rate $R$ between different realisations.
We assume this variation between
realisations is a reasonable model for variations between species.
Returning to eq.~(\ref{hetnuc}) for the rate, using
eq.~(\ref{ri2})  we obtain
\begin{equation}
R=N_s^{-1}\nu\exp\left(-\Delta F^*_0\right)
\sum_i^{N_s} \exp\left(n_i\epsilon\right),
\label{hetnuc2}
\end{equation}
where the $n_i$ are taken to be random variables
drawn from the Gaussian distribution eq.~(\ref{pdfni}).
Except for constant factors, the rate $R$ is equivalent to
the partition function of the Random Energy Model (REM) of
Derrida \cite{derrida80,note}. The REM is a simple and well
studied model of glasses and other disordered systems.

Just as the average partition function of the REM can be obtained,
we can obtain the average of the rate $R$,
\begin{eqnarray}
\langle R\rangle&=&N_s^{-1}\nu\exp\left(-\Delta F^*_0\right)
\langle\sum_{i=1}^{N_s} \exp\left(n_i\epsilon\right)\rangle\\
&= &\nu\exp\left(-\Delta F^*_0\right)
\exp\left(m\epsilon+\epsilon^2w^2/2\right).
\label{nuc2}
\end{eqnarray}
This is the average of $R$ over many different realisations of the surface,
i.e., many different sets of the $N_s$ random variables $n_i$ that
define a surface. As $R$ is a sum over random variables it itself
is a random variable.
For the large values of $N_s$ considered here, the rate
$R$ is either self-averaging or non-self-averaging. It
is self-averaging if for almost all realisations
the rate of unfolding
$R$ is close to $\langle R\rangle$, i.e.,
if $R$ is almost the same for almost all realisations.
Then the right-hand side
of eq.~(\ref{nuc2}) will be a good approximation to the rate $R$
of any realisation. If it is non-self-averaging then the rate $R$
differs appreciably from one realisation to another, the values
of $R$ have a large spread and eq.~(\ref{nuc2}) is unlikely to provide
a good approximation to the value of $R$ for a randomly selected
realisation. $R$ is non-self-averaging if and only if the sum
of eq.~(\ref{hetnuc2}) is dominated by one or a few terms:
the variation comes from variation in the values of the
largest terms in the sum. This is just as in the REM, see
ref.~\cite{derrida80} for details.


Recall that we are assuming that a realisation corresponds to a species.
Thus, if $R$ is self-averaging, then our model
predicts that the rate of unfolding of a particular protein is
almost the same in all or almost all species, whereas if it is not
self-averaging then the rate of unfolding of a specific
protein will vary significantly
from one species to another.

We will now determine the boundary where the rate $R$
crosses over from self-averaging to non-self-averaging.
From eq.~(\ref{hetnuc2})
we see that the rate $R$ is dominated by unfolding configurations
with values of $n_i$ where
the product of the number of configurations
and $\exp(n_i\epsilon)$, is a maximum. The number of configurations
is simply proportional to the probability of eq.~(\ref{pdfni}).
The maximum of the product $p(n_i)\exp(n_i\epsilon)$ is at
$n_{max}=m+\epsilon w^2.$
Now, the {\em average} number of configurations around this value of $n_i$
is just $N_sp(n_{max})$, and because this average is a sum
over independent random variables (the $n_i$) the ratio
of the fluctuations to the mean scales as $[N_sp(n_{max})]^{-1/2}$.
Thus the fluctuations in the number of configurations that contribute the
dominant amount to the rate, and hence the fluctuations in the rate
itself are small relative to the mean if and only if
$N_sp(n_{max})\gg 1$. This is true whenever $2\ln N_s-\epsilon^2w^2>0$.

Thus, the boundary between self-averaging and non-self-averaging regimes
is given by the equation
\begin{equation}
2\ln N_s-\epsilon^2w^2=0.
\label{transit}
\end{equation}
Note that $\epsilon^2w^2$ is the variance of the distribution
of interaction energies between the transition state and the surface.
Thus the rate is self-averaging if and only if the logarithm of the
number of possible configurations that the transition state can unfold in,
is larger than half the variance of the interaction energy. This
is the main result of this work. It is a very general result ---
it applies generally to activated processes in a random or near-random
environment. Our conclusions
here apply to any process
with a rate given by an equation of the form of
eq.~(\ref{hetnuc}), not just to protein unfolding {\it in vivo}.
See ref.~\cite{sear04} for an application to nucleation at
first-order phase transitions.

In the non-self-averaging regime, a single unfolding configuration can
be responsible
for a significant fraction of the entire rate of unfolding at the
surface. This configuration must of course be the
configuration with the largest
value of $n_i$. We denote this largest
value by $x$.
If we define the
probability distribution function, $p_{ev}(x)$, of $x$, then
the fraction of the rate $R$ that is due to this extreme
value is,
\begin{equation}
f_{ev}=\frac{\nu\exp\left(-\Delta F_0^*\right)}{N_s\langle R\rangle}
\int p_{ev}(x)\exp\left(\epsilon x\right){\rm d}x.
\label{fev1}
\end{equation}
We can simplify eq.~(\ref{fev1}) by introducing the reduced
variable $y=(x-m)/w$. Then, from eq.~(\ref{fev1}) and using
eq. (\ref{nuc2}) for $\langle R\rangle$,
we obtain
\begin{equation}
f_{ev}=N_s^{-1}\exp\left(-(\epsilon w)^2/2\right)
\int {\rm d}y\exp\left(\epsilon wy\right)p_{ev}(y),
\label{fev2}
\end{equation}
where $p_{ev}(y)$ is the probability distribution function for
the maximum value of a set of $N_s$ values taken from a Gaussian
of zero mean and unit standard deviation.
Note that although the absolute value of the rate $R$ and
of the contribution of the extreme value both depend on the mean $m$,
$f_{ev}$ does not. It depends
only on the product $\epsilon w$, and on $N_s$.

The determination of $p_{ev}(y)$ is a standard problem in extreme-value
statistics \cite{sornette}.
We start from the fact that
the probability that the largest of $N_s$ values is $y$
is the probability that 1 of the $N_s$ configurations has
a value $y$, and all
the remaining $N_s-1$ configurations have smaller values,
multiplied by $N_s$, as
any one of the $N_s$ configurations can have the largest value. Thus,
\begin{equation}
p_{ev}(y)=N_sp(y)p_{<}^{N_s-1}(y),
\label{pev1}
\end{equation}
where $p(y)$ is a normalised Gaussian of zero mean and unit
standard deviation, and
$p_{<}(y)$ ($p_{>}(y)$) is the probability of obtaining a number
less (greater) than
$y$ from a Gaussian of zero mean and unit standard deviation.
We are interested in the region where $x$ is several standard
deviations above the mean, $y\gg 1$.
Now, $p_{<}=1-p_{>}$, and so
for $y\gg 1$, $p_{>}\ll 1$, and we can rewrite
eq.~(\ref{pev1}) as
\begin{equation}
p_{ev}(y)\simeq N_sp(y)\exp\left[-N_sp_{>}(y)\right].
\label{pev2}
\end{equation}
Also, $p_{>}(y)=(1/2)\mbox{erfc}(y/2^{1/2})$, which
for $y\gg 1$ simplifies to
$
p_{>}(y)\simeq \exp\left(-y^2\right)/[
(2\pi)^{1/2}y].
$

\begin{figure}
\caption{
\lineskip 2pt
\lineskiplimit 2pt
The mean fraction, $f_{ev}$, of the rate $R$ that is due to the configuration
with the largest $n_i$,
as a function
of the product of the width of the Gaussian, $w$, and the interaction
energy $\epsilon$.
The solid, dashed and dotted curves are
for $N_s=$1,000, 10,000 and 100,000 configurations, respectively.
\label{figfev}
}
\vspace*{0.5cm}
\begin{center}
\epsfig{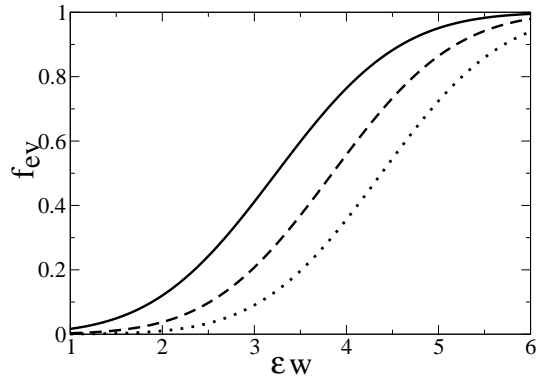}
\vspace*{-0.5cm}
\end{center}
\end{figure}

In fig.~\ref{figfev} we have plotted the fraction of the rate due
to the configuration with the largest interaction energy, and so the lowest
barrier, $f_{ev}$, as a function of $\epsilon w$.
We took $N_s=$1,000, 10,000 and 100,000.
Assuming that there are a few thousand different
species inside a cell and that each can potentially interact
with the transition state in a few ways, we end up with the estimate
$N_s\approx 10^4$. The other parameter is $\epsilon w$. The
interaction strength of a pair of monomers is expected to lie
in the range 1 to 3 (recall that $\epsilon$ is in units of $k_BT$).
If the fraction of hydrophobic monomers $h\approx 1/2$, then
for $B\approx 5$ to 15 hydrophobic monomers, we have that
$w\approx 1$ to 2. Combining these values for $\epsilon$ and $w$
we have that $\epsilon w\approx 0.5$ to 6.
Returning to fig.~\ref{figfev}
we see that as $\epsilon w$ increases, so does $f_{ev}$.
For $N_s=$10,000, eq.~(\ref{transit}) is satisfied for
$\epsilon w=4.29$. For $\epsilon w$ around this value the configuration
with the largest interaction energy already contributes a large
amount to the total rate, on average. This large contribution
will vary significantly from one realisation to the next, from one
species to the next. So, the rate of unfolding of the protein will
vary significantly from one species to the next, depending on whether
the species has some part of a protein, RNA molecule etc., that binds to the
transition state unusually strongly. Our estimate for the
possible values of $\epsilon w$ {\it in vivo}
goes up to around 6, so we estimate
that the variation in the interaction free energies with a transition
state may be large enough to cause random species-to-species
variation.
The RNA molecule or molecules found to catalyse the conversion
is within our model the origin of one of the configurations
that are outliers of the distribution, that interact most
strongly with the transition state.
Of course if $\epsilon w$ is small then the rate $R$ has significant
contributions from many unfolding configurations and so varies weakly
from species to species, essentially due to variations in the rate being
averaged out in accordance with the central-limit theorem.


In conclusion,
Supattapone and coworkers \cite{deleault03} have found
that cell extracts
of some species but not others accelerate
the conversion of the prion protein to a protease-resistant form.
This conformational change must involve partial unfolding.
Protein unfolding {\it in vivo} or in a cell extract
occurs in a very complex and heterogeneous
environment. There are a huge number of species present
that potentially could interact with and stabilise the transition
state of unfolding. A single strongly stabilising interaction
could dramatically increase the rate of unfolding.
Here we have suggested a possible
model for
the species-to-species variation in the ability of cell extracts
to accelerate prion protein conversion
\cite{deleault03}.
The model is a statistical one:
interactions are modelled by random variables
and different species by different uncorrelated
realisations of the random interactions.
We suggest that the
acceleration is due to a strong interaction of the transition
state for prion protein conversion
with one or a few species of RNA molecules,
and that this interaction is strong
simply by chance.
It is simply accidental that they
reduce the free-energy barrier to unfolding. 
Proving this suggestion would require identifying the RNA
molecule or molecules that interact with the prion protein
and then demonstrating that there is no functional relationship
between the protein and the RNA. Falsifying the suggestion is
perhaps more straightforward, it only requires finding
a functional relationship.
The species-to-species variation then simply comes
from the variation in the nucleotide sequences
of RNA molecules from species to species. The RNA
molecules that perform the same function in say mice and fruit flies,
will have similar but not identical nucleotide sequences
and so will have different interaction free energies
with the transition state.
Finally, it should be noted that
it is also possible that the RNA molecule or molecules
have evolved to interact with the prion protein, although
we know of no evidence that they are under selection pressure
to interact specifically with the transition state.

It is a pleasure to acknowledge that this work has benefited
greatly from discussions with J. Cuesta.
This work was supported by The Wellcome Trust (069242).



\begin{thebibliography}{99}

\bibitem{dobson01} DOBSON C.~M.,
Phil. Trans. Royal Soc. B, 356 (2001) 133

\bibitem{harris99} HARRIS D.~A.,
Clin. Micro. Rev., 12 (1999) 429

\bibitem{aguzzi04} AGUZZI A. and POLYMENIDOU M.,
Cell, 116 (2004) 313

\bibitem{finkelstein} FINKELSTEIN A.~V. and PTITSYN O.~G.,
Protein Physics
(Academic Press, London) 2002

\bibitem{creighton} CREIGHTON T.~E.,
Proteins: Structures and Molecular Properties
(Freeman, New York) 1993

\bibitem{deleault03} DELEAULT N.~R., LUCASSEN R.~W. and SUPATTAPONE S.,
Nature, 425 (2003) 717

\bibitem{cordeiro01} CORDIERO Y., MACHADO F., JULIANO L.,
APARECIDA JULIANO M., BRENTANI R.~R., FOGUEL D. and SILVA J.~L.,
J. Bio. Chem., 276 (2001) 49400

\bibitem{ryou03} RYOU C., PRUSINER S.~B. and LEGNAME G.,
J. Mol. Bio., 329 (2003) 323

\bibitem{foldnote} Even {\it in vitro}
protein unfolding is in general more complex
than a simple crossing of a single barrier
\cite{finkelstein,onuchic97,pande00}. Also, it is far from
clear that protein unfolding is the rate limiting step in
the progression of the formation of the protein fibrils in
prion diseases. Here we simply assume that it is a simple
activated process and explore the consequences.

\bibitem{wigner67} WIGNER E.,
SIAM Rev., 9 (1967) 1

\bibitem{sear03} SEAR R.~P. and CUESTA J.~A.,
Phys. Rev. Lett., 91 (2003) 245701

\bibitem{karpov96} KARPOV V.~G. and OXTOBY D.~W.,
Phys. Rev. B, 54 (1996) 9734

\bibitem{sear04} SEAR R.~P.,
Phys. Rev. E, {\bf 70}, 021605 (2004); cond-mat/0406019.

\bibitem{derrida80} DERRIDA B., Phys. Rev. Lett., 45 (1980) 79;
Phys. Rev. B, 24 (1981) 2613

\bibitem{note} The REM is used extensively in the study of
protein folding, see for example refs.~\cite{onuchic97,pande00}.
But our use of the REM is rather different, the random variables
here are interaction energies between the transition state and the surface,
whereas in studies of protein folding they are the states of an isolated
protein.

\bibitem{sornette} SORNETTE D.,
Critical Phenomena in Natural Sciences
(Springer-Verlag, Berlin) 2000

\bibitem{onuchic97} ONUCHIC J.~N., LUTHEY-SCHULTEN Z. and WOLYNES P.~G.,
Ann. Rev. Phys. Chem., 48 (1997) 545

\bibitem{pande00} PANDE V.~S., GROSBERG A.~Yu. and TANAKA T.,
Rev. Mod. Phys., 72 (2000) 259











\end{thebibliography}
\end{document}